\documentclass[]{iopart}
\usepackage{graphicx, epsfig} %include figure files
\usepackage{bm}

\expandafter\let\csname equation*\endcsname\relax
\expandafter\let\csname endequation*\endcsname\relax

\usepackage[caption=false]{subfig}
\usepackage[usenames]{xcolor}

\usepackage{marvosym}
\usepackage{yfonts}
\usepackage{upgreek}
\usepackage{pdflscape}
\definecolor{linkcolor}{rgb}{0.0, 0.3, 0.5}
\definecolor{purple}{rgb}{0.7, 0.05, 0.5}
\usepackage[unicode,colorlinks=true,citecolor=linkcolor,linkcolor=linkcolor,urlcolor=linkcolor]{hyperref}
\usepackage[all]{hypcap}
\usepackage{amsmath,amsfonts,amssymb}
\usepackage{bm}
\usepackage{longtable}
\usepackage[T1]{fontenc}
\usepackage[utf8]{inputenc}
\usepackage{verbatim}
\usepackage{pifont}
\usepackage{yfonts}
\usepackage{url}
\usepackage{multirow}
\usepackage{tensor}
\usepackage{mathtools}
\usepackage[normalem]{ulem}
\usepackage{fancyhdr}
\usepackage{xcolor}
\usepackage{soul}

\renewcommand{\vec}[1]{\boldsymbol{#1}}

\begin{document}
\newcounter{count}

\pagestyle{fancy}\lhead{Spontaneous scalarisation with self-interacting potentials}
\chead{}
\rhead{\thepage}
\lfoot{}
\cfoot{}
\rfoot{}

\begin{center}
\title{\large Inverse-chirp signals and spontaneous scalarisation with self-interacting potentials in stellar collapse}
\end{center}

\author{
Roxana Rosca-Mead$^{1}$,
Christopher J Moore$^{2}$,
Michalis Agathos$^{3}$,
Ulrich Sperhake$^{1,4}$
}

\address{$^{1}$~Department of Applied Mathematics and Theoretical Physics,
Centre for Mathematical Sciences, University of Cambridge,
Wilberforce Road, Cambridge CB3 0WA, United Kingdom}

\address{$^{2}$~Institute for Gravitational Wave Astronomy and School of Physics and Astronomy, University of Birmingham, Edgbaston, Birmingham B15 2TT, UK}

\address{$^{3}$~Theoretical Physics Institute, University of Jena, 07743 Jena, Germany}

\address{$^{4}$~Theoretical Astrophysics 350-17,
California Institute of Technology,
1200 E California Boulevard, Pasadena, CA 91125, USA}

\ead{rr417@cam.ac.uk}

\begin{abstract}
We study how the gravitational wave signal from stellar collapse in
scalar-tensor gravity varies under the influence
of scalar self-interaction.
To this end, we extract the gravitational radiation
from numerical simulations of stellar collapse for a range of potentials
with higher-order terms in addition to the quadratic mass term.
Our study includes collapse to neutron stars and black holes and we find the strong inverse-chirp signals obtained for the purely quadratic potential to be exceptionally robust under changes in the
potential at higher orders;
quartic and sextic terms in the potential lead to
noticeable differences in the wave signal only if their contribution is
amplified, implying a relative fine-tuning to within 5 or more orders
of magnitude between the mass and self-interaction parameters.
\end{abstract}

\maketitle

%=============================================================================
\section{Introduction}
\label{sec:intro}
In spite of the tremendous success of general relativity (GR) in
explaining a plethora of phenomena in the observable universe
\cite{Psaltis:2008bb,Will:2014kxa}, theoretical considerations
as well as persistent puzzles in observational astronomy may ultimately require
Einstein's theory to be extended or modified
\cite{Berti:2015itd}. The new era of gravitational wave (GW)  astronomy, marked by LIGO's detection of GW150914
\cite{Abbott:2016blz} and
ten further events identified since~\cite{LIGOScientific:2018mvr}, in the data of the LIGO~\cite{TheLIGOScientific:2014jea}
and Virgo~\cite{TheVirgo:2014hva} detectors, has now opened
up qualitatively new opportunities to probe GR
in the strong-field regime~\cite{LIGOScientific:2019fpa}. First tests
using GW observations have resulted in
valuable constraints on deviations from GR, including constraints on
the post-Newtonian coefficients and on the propagation of the dipole radiation
from compact binaries \cite{Yunes:2016jcc,TheLIGOScientific:2016src,Abbott:2018lct}, but more
comprehensive tests will require more systematic efforts
on the modelling of sources of GWs in
modified theories of gravity.

This effort faces considerable challenges. First,
there should exist a well-posed initial value formulation
of the theory under consideration to ensure that solutions
are unique and depend continuously on the initial data. This aspect has been explored for a few
candidate theories \cite{Delsate:2014hba,Papallo:2017qvl,Allwright:2018rut,Papallo:2019lrl}
but largely remains an open issue
which is further complicated by the fact that the question of well-posedness is, in general, gauge dependent.
The difficulties associated with the potential lack
of well-posedness may be bypassed through
a perturbative expansion of the theory around
GR and truncating this series at some order
in the expansion parameter. Such an effective-field theory
approach has been used to compute deviations at linear
order from GR
in the inspiral of black-hole (BH) binaries in
dynamical Chern-Simons theory \cite{Okounkova:2017yby} and scalar Gauss-Bonnet
gravity \cite{Witek:2018dmd}.
A second major challenge faced in the modelling of
sources of GWs in modified gravity arises from the
tremendous success of GR itself. Clearly, a serious
candidate theory
must be compatible with the wide range
of observational tests GR has already passed.
The challenge then is to identify theories that
agree with GR in the weak-field regime but make
concrete predictions in the strong-field regime that
deviate sufficiently from GR such that they serve
as a potential discriminant in observational tests
\cite{Berti:2015itd}.

The most concrete prediction of this type is the
{\em spontaneous scalarisation} of compact stars
in scalar tensor (ST) theory identified by Damour and
Esposito-Far{\`e}se \cite{Damour:1993hw}. Here, a second
branch of strongly scalarised stars emerges over a
significant subset of the parameter space of the theory
and its members may be energetically favoured over their
weakly scalarised, ``GR like'' counterparts. The onset
of this strong scalarisation has been observed in various
simulations of collapsing stars and
binary neutron star mergers in ST
theory where it generates significant gravitational radiation
in the form of scalar waves 
\cite{Scheel:1994yr,Harada:1996wt,Novak:1997hw,Novak:1998rk,Novak:1999jg,Palenzuela:2013hsa,Palenzuela:2015ima,Gerosa:2016fri}. All of these
studies, however, as well as recent analytical calculations
of binary dynamics in ST theory to high post-Newtonian order~\cite{Bernard:2018hta,Bernard:2018ivi},
are restricted to the case of {\em massless}
ST gravity.
In the massless case, the parameter space of the
theory is already severely constrained through binary
pulsar observations
\cite{Freire:2012mg,Antoniadis:2013pzd,Wex:2014nva}
and 
Doppler tracking of the Cassini spacecraft
\cite{Bertotti:2003rm}. The surviving parameter regime
of massless ST gravity barely allows for
spontaneous scalarisation to occur.

On the other hand, both the binary pulsar and Cassini constraints
rely on observations of widely separated objects and therefore
do not apply to theories where the scalar degree of freedom
is effectively screened on the scales in question.
Such a screening is provided by a scalar mass
$\mu \gtrsim 10^{-19}\,\mathrm{eV}$ corresponding to a
Compton wavelength $\lambda_c=(2\pi \hbar)/(\mu c)$ smaller
than or comparable to the distances between the relevant
objects \cite{Alsing:2011er,Ramazanoglu:2016kul}.

Massive ST gravity therefore remains a largely
untested class of theories with considerable potential
to generate strong-field deviations from GR in, as yet,
unobserved regimes while passing all weak-field tests. The
theory is furthermore manifestly well-posed by virtue of the
equivalence of its description in the Einstein and
Jordan frames
\cite{Damour:1992we,Damour:1996ke,Salgado:2005hx,Salgado:2008xh}.
Recent years have accordingly seen a rising number of studies
exploring potentially observable features of compact
objects in this class of theories,
such as the computation of static equilibrium
models \cite{Ramazanoglu:2016kul}, the structure of
uniformly or differentially rotating neutron stars
and their inertia and quadrupole moment
\cite{Doneva:2016xmf,Yazadjiev:2016pcb,Doneva:2018ouu}
and the generalisation of the
spontaneous scalarisation mechanism to
a wider range of theories
\cite{Horbatsch:2015bua,Ramazanoglu:2017xbl,Ramazanoglu:2019gbz}.

In previous work \cite{Sperhake:2017itk}, we have computed
the first GW signals in massive ST
gravity and demonstrated how the gravitational collapse of
stellar cores leads to highly characteristic signals,
stretched out over years or even centuries by the dispersive
nature of the mass term, that would show up in
existing LIGO-Virgo searches. Most recently, a first exploration
of stellar collapse in ST theory with a mass {\em and}
a quartic self-interaction term has identified the possibility
of a weakening effect of the self-interaction term on the
magnitude of the GW signal\,\cite{Cheong:2018gzn}.
The main purpose of this study is to perform a systematic
exploration of the impact of higher-order terms in the
scalar potential and test the robustness of the
{\em inverse-chirp} signals found in \cite{Sperhake:2017itk}.
As we will demonstrate in the
remainder of this paper, the generation of the signals, their
amplitude and propagation is
astonishingly robust to modifications of the potential.

This study is organised as follows. In Sec.~\ref{sec:formalism}
we review the formalism of our simulations, with a focus on the
modifications in the scalar potential. The generation of
GW signals in stellar collapse is modelled and compared
to the case of non-interacting scalars in
Sec.~\ref{sec:results}. In Sec.~\ref{sec:propagation}, we
discuss the propagation of the signal from radii
comparable to our computational domain to large distances
in the far-field region. Throughout this work, we use
natural units where the speed of light and gravitational
constant $c=G=1$.

%=============================================================================
\section{Formalism}
\label{sec:formalism}
We consider in this work the class of ST
theories of gravity first studied by
Bergmann \cite{Bergmann:1968ve} and Wagoner
\cite{Wagoner:1970vr} which are characterised by
the following properties. The theory is described
by an action $S=S_{\rm G}+S_{\rm M}$, where $S_{\rm G}$
only contains the gravitational fields and $S_{\rm M}$
describes the matter fields and their interaction with gravity.
Gravity is mediated by the (physical)
spacetime metric $g_{\alpha\beta}$
and a single, non-minimally coupled real scalar field.
Variation of the action $S$ results in at most two-derivative
field equations\footnote{The word ``two-derivative'' implies
that each term may involve at most two derivative operators,
i.e.~terms may be linear in second derivatives or quadratic
in first derivatives but may not contain products such as
$f_{,\alpha}\, f_{,\beta\gamma}$.}
that are diffeomorphism
invariant and obey the weak equivalence principle. This
class of theories is conveniently described in the
so-called Einstein frame, obtained from the physical or Jordan
frame through a conformal transformation of the metric
and a redefinition of the scalar degree of freedom. In
natural units,
the general action for these theories is given by
\cite{Fujii:2003pa,Berti:2015itd}
\begin{equation}
  S=\int {\rm d} x^4 \frac{\sqrt{-\bar{g}}}{16}
  \left[ \bar{R}-2\bar{g}^{\alpha \beta}\partial_{\alpha}\varphi \partial_{\beta}\varphi-4V(\varphi)\right]
  +S_{\rm M}(g_{\alpha\beta},\psi_{\rm M})\,,
\end{equation}
where $\varphi$ is the scalar field,
$\psi_{\rm M}$ collectively denotes all matter fields,
and $\bar{R}$ and
$\bar{g}$ are the Ricci scalar and metric determinant constructed
from the conformal metric
\begin{equation}
  \bar{g}_{\alpha \beta} = F(\varphi) g_{\alpha \beta}\,.
\end{equation}
In the Einstein frame, the scalar
field is minimally coupled to the conformal metric
$\bar{g}_{\alpha \beta}$ but the matter fields couple to the
{\em physical} or {\em Jordan-Fierz} metric $g_{\alpha\beta}$.
This class of theories has two free functions, the scalar
potential $V(\varphi)$ and the conformal factor
$F(\varphi)$; we will discuss both in more detail below.

Our study concerns the gravitational collapse of
stellar cores at the end of their nuclear burning phase and
the GW signal generated in the rapid
transition of a low-density star to a high-density compact object. The
main dynamics of this process are dominated by the sudden
radial compression of the stellar matter and we therefore
employ spherical symmetry in this study. More specifically,
we use radial gauge and polar slicing in the Einstein frame
and write the conformal metric as
\begin{equation}
  \bar{g}_{\alpha \beta}\mathrm{d}x^{\alpha} \mathrm{d}x^{\beta} =
  -F\alpha^2 \mathrm{d}t^2+FX^2\mathrm{d}r^2
  + r^2 (\mathrm{d}\theta^2+\sin^2\theta\mathrm{d}\phi^2)\,,
\end{equation}
where $\alpha$ and $X$ are functions of $(t,r)$. We model
the stellar matter as a perfect fluid with baryon density
$\rho$, pressure $P$, internal energy $\epsilon$,
and 4-velocity $u^{\alpha}=(1-v^2)^{-1/2}
[\alpha^{-1},\,vX^{-1},\,0,\,0]$. Here, $\rho$, $P$, $\epsilon$
and $v$ are also functions
of $(t,r)$, and the energy momentum tensor is given by
\begin{equation}
  T_{\alpha\beta}=(\rho + \rho \epsilon + P)u_{\alpha}
  u_{\beta} + Pg_{\alpha\beta}\,.
\end{equation}
Following Refs.~\cite{Gerosa:2016fri,Sperhake:2017itk},
we use a hybrid equation of state (EOS) where the pressure
consists of a cold and a thermal part, $P=P_{\rm c}+P_{\rm th}$,
defined in terms of the cold contribution to the internal energy
$\epsilon_c$ by
\begin{eqnarray}
  &&\rho \le \rho_{\rm nuc}\,:~~
  P_{\rm c} = K_1 \rho^{\Gamma_1}\,,~~~~~
  \epsilon_{\rm c} =
  \frac{K_1}{\Gamma_1 -1}\rho^{\Gamma_1-1}\,,
  \nonumber \\[5pt]
  &&\rho > \rho_{\rm nuc}\,:~~
  P_{\rm c} = K_2 \rho^{\Gamma_2}\,,~~~~~
  \epsilon_{\rm c}=\frac{K_2}{\Gamma_2-1}
  \rho^{\Gamma_2-1}+E_3
    \label{eq:Pc}\,, \\[10pt]
  &&P_{\rm th} = (\Gamma_{\rm th}-1)\,\rho\,(\epsilon-
  \epsilon_{\rm_{\rm c}})\,.
  \label{eq:Pth}
\end{eqnarray}
Here we use a nuclear density
$\rho_{\rm nuc}=2\times 10^{14}~{\rm g\, cm}^{-3}$,
$K_1=4.9345\times 10^{14}\,{\rm [cgs]}$ \cite{Shapiro1983},
and $K_2$ and $E_3$ follow from continuity across 
$\rho_{\rm nuc}$. There thus remain 3 free parameters
determining the EOS, the polytropic exponents $\Gamma_1$,
$\Gamma_2$ and the thermal coefficient $\Gamma_{\rm th}$.
In the remainder of this work we set these to the fiducial
values $\Gamma_1=1.3$, $\Gamma_2=2.5$, $\Gamma_{\rm th}=1.35$
of Ref.~\cite{Gerosa:2016fri}; see Sec.~3.1 and references
therein for further details.
By performing additional
simulations, we have found that the impact of the self-interaction
terms is not significantly affected by varying the EOS.

The field and matter equations determining the time evolution
of an initial stellar profile with the above EOS in
ST theory and spherical symmetry are then given by
Eqs.~(3)-(6) in Ref.~\cite{Sperhake:2017itk}. The key difference
to that study is that we now consider a wider class of
scalar potentials,
\begin{equation} \label{eq:Def_Potential}
  V(\varphi)=\frac{\mu^2 \varphi^2}{2\hbar^2}\left(
  1+\lambda_1 \frac{\varphi^2}{2} + \lambda_2
  \frac{\varphi^4}{3}+ \ldots+
  \lambda_n \frac{\varphi^{2n}}{n+1} \right)\,,~~~
  \text{with}~~\lambda_n > 0\,.
\end{equation}
Here, the scalar mass $\mu$ introduces a characteristic frequency
\begin{equation}
  \omega_* = 2\pi f_* = \frac{\mu}{\hbar}\,.
\end{equation}
In this work we set $\mu=10^{-14}\,{\rm eV}$ corresponding to
$\omega_* = 15.2\,{\rm s}^{-1}$ or $f_*=2.42\,{\rm Hz}$.
Note that all $\lambda_i$ in Eq.~(\ref{eq:Def_Potential}) are dimensionless and that we
recover the massive but not self-interacting
case of \cite{Sperhake:2017itk} by setting $\lambda_i=0$, and the massless case by setting $\mu=0$.

The remaining free function in our theory is the
conformal factor or coupling function $F(\varphi)$.
The standard choice to parametrize this function,
\footnote{A common alternative notation for the conformal factor
is $A(\varphi):=F(\varphi)^{-1/2}$.}
\begin{equation}
  F(\varphi) = e^{-2\alpha_0 \varphi-\beta_0 \varphi^2}\,,
  \label{eq:F}
\end{equation}
is motivated by the fact that in this form, $\alpha_0$
and $\beta_0$ completely determine all modifications of
gravity at first post-Newtonian order
\cite{Damour:1992we,Damour:1996ke,Chiba:1997ms}. In the
remainder of our study we will follow this choice and
work with the conformal factor in Eq.~(\ref{eq:F}).

The parameter $\alpha_0$ is related to the logarithmic
derivative of the conformal factor,
\begin{equation}
  \alpha_0 = -\frac{1}{2} \left. \frac{\mathrm{d} \ln F}{\mathrm{d}\varphi}
  \right|_{\varphi = 0}\,,
\end{equation}
and determines the coupling of the scalar field to a GW detector
far away from the source: A scalar wave of angular frequency
$\Omega$ induces a total detector strain
\begin{equation}
  h = 2\alpha_0 \left[1 -\left(\frac{\omega_*}{\Omega}\right)
    \right]\varphi\,.
\end{equation}
For massless fields, this reduces to the familiar $h=2\alpha_0
\varphi$, but for a massive scalar, the strain is mildly
reduced because the scalar mode splits into a transverse and
a longitudinal component; see \cite{Sperhake:2017itk} for more
details.

Let us conclude this section with a summary of the
free parameters in our ST theory.
The conformal factor is
described by $\alpha_0$ and $\beta_0$ and we have $n$
further parameters describing the self-interaction of the
scalar field. In this paper we restrict our attention to the case $n=2$ giving a total of four free parameters,
$\alpha_0,~\beta_0,~\lambda_1,~\lambda_2,$
additionally to the four fixed parameters $\mu=10^{-14}\,{\rm eV}$,
$\Gamma_1=1.3$, $\Gamma_2=2.5$,
$\Gamma_{\rm th}=1.35$ for the scalar mass and the EOS.

%=============================================================================
\section{Models and results}
\label{sec:results}

The computational framework for our core collapse simulations
is based on the open source code GR1D \cite{O'Connor:2009vw} for
modelling spherically symmetric fluids in GR with high-resolution
shock capturing schemes. In Refs.~\cite{Gerosa:2016fri}
and \cite{Sperhake:2017itk}, GR1D has been extended to ST gravity
with a massless or a massive scalar field, respectively. Here
we employ the same setup as used in \cite{Sperhake:2017itk}
except that the scalar potential is now given by
Eq.~(\ref{eq:Def_Potential}).
The runs presented in this paper have been performed with an inner uniform grid of $\Delta r=250 $ m up to $r=40$ km and an outer logarithmic grid up to $r=9\times10^5$ km with a total of
$10\,000$ grid points. At this resolution, the waveforms
incur a discretization error of about $6\,\%$ 
\cite{Sperhake:2017itk}.
 
The initial data for our simulations are given
by the catalogue of non-rotating pre-super-nova stellar profiles
of Woosley and Heger \cite{Woosley:2007as,WHsite} with the
scalar field and its time derivative initialised to zero.
This catalogue covers progenitor models with
zero-age-main-sequence (ZAMS) masses ranging between $(12-75)\,M_{\odot}$
and different metallicity. Here we focus on two stellar
models with $10^{-4}$ times solar metalicity and
$M_{\rm ZAMS}=39\,M_{\odot}$ and
$41\,M_{\odot}$. We have studied a wider range of initial models
and also tested the effect of varying the EOS parameters; the impact
of the self-interaction terms $\lambda_i$ on the
GW generation exhibits a universal character in all these
simulations which is fully encapsulated by the models
presented in this section.

\begin{figure}[t]
	\centering
	\label{fig:vary_lambda1}
	\includegraphics[width=1.0\textwidth]{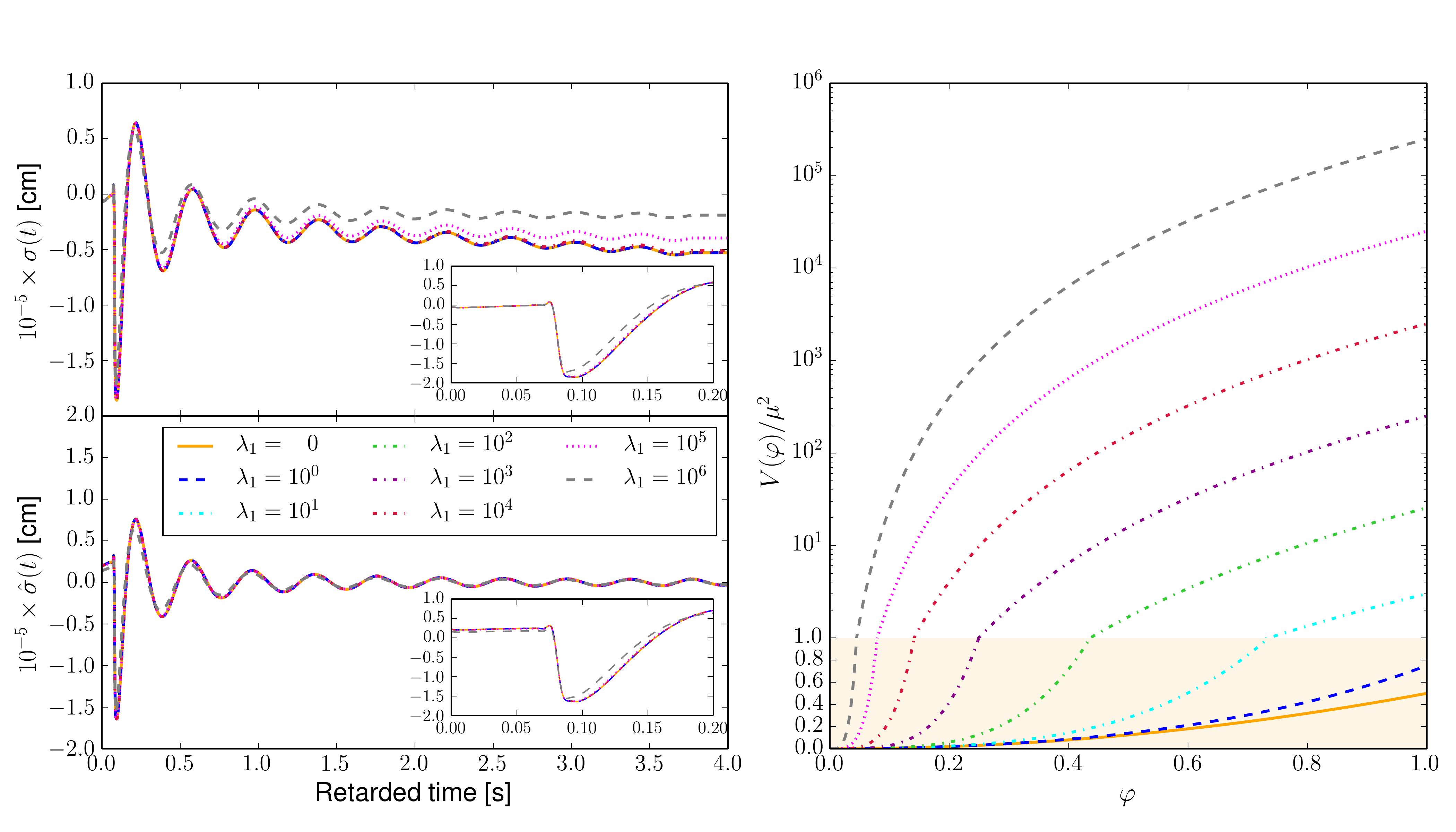}
	\caption{The effect of varying the quartic, $\lambda_1$, term of the scalar field potential. Top-left panel: the scalar field signal at the extraction radius $7\times 10^4$ km. Bottom-left panel: the band-passed scalar field signal at the extraction radius $7\times 10^4$ km; modes with frequencies less than $\omega_*$ are set to $0$. Right-hand panel: the scalar field potential. The progenitor used is a $ 41 \ M_{\odot}$ star of $10^{-4}$ solar metallicity  and the scalar parameters are $\alpha_0=10^{-2}$, $\beta_0=-20$.
	%, $\mu=10^{-14}$ eV.
	}
\end{figure}

The one common feature of
all our simulations is that they result in strong
scalarisation of the compact stars formed during
the collapse and, thus, lead to a large GW signal
for ST theory without self-interaction, i.e.~for
$\lambda_i=0$. Let us first consider
the collapse of a 41~$M_{\odot}$ progenitor model
with $10^{-4}$ solar metallicity with ST parameters $\alpha_0=10^{-2}$, $\beta_0=-20$.
For $\lambda_i=0$, the collapse of the baryon matter
leads to core bounce at $t\simeq0.07$~s
which generates a scalar wave signal $\sigma=r_{\rm ex}\varphi$ of
$O\left(10^5\right)$ at extraction radius
$r_{\rm ex}=7\times 10^4~{\rm km}$. This signal
is shown as the solid (orange) curve
in Fig.~\ref{fig:vary_lambda1}. During the collapse,
the scalar field at the centre of the star rapidly increases
before levelling off at a magnitude of $0.38$.

\begin{figure}[t]
	\centering
	\includegraphics[width=1.0\textwidth]{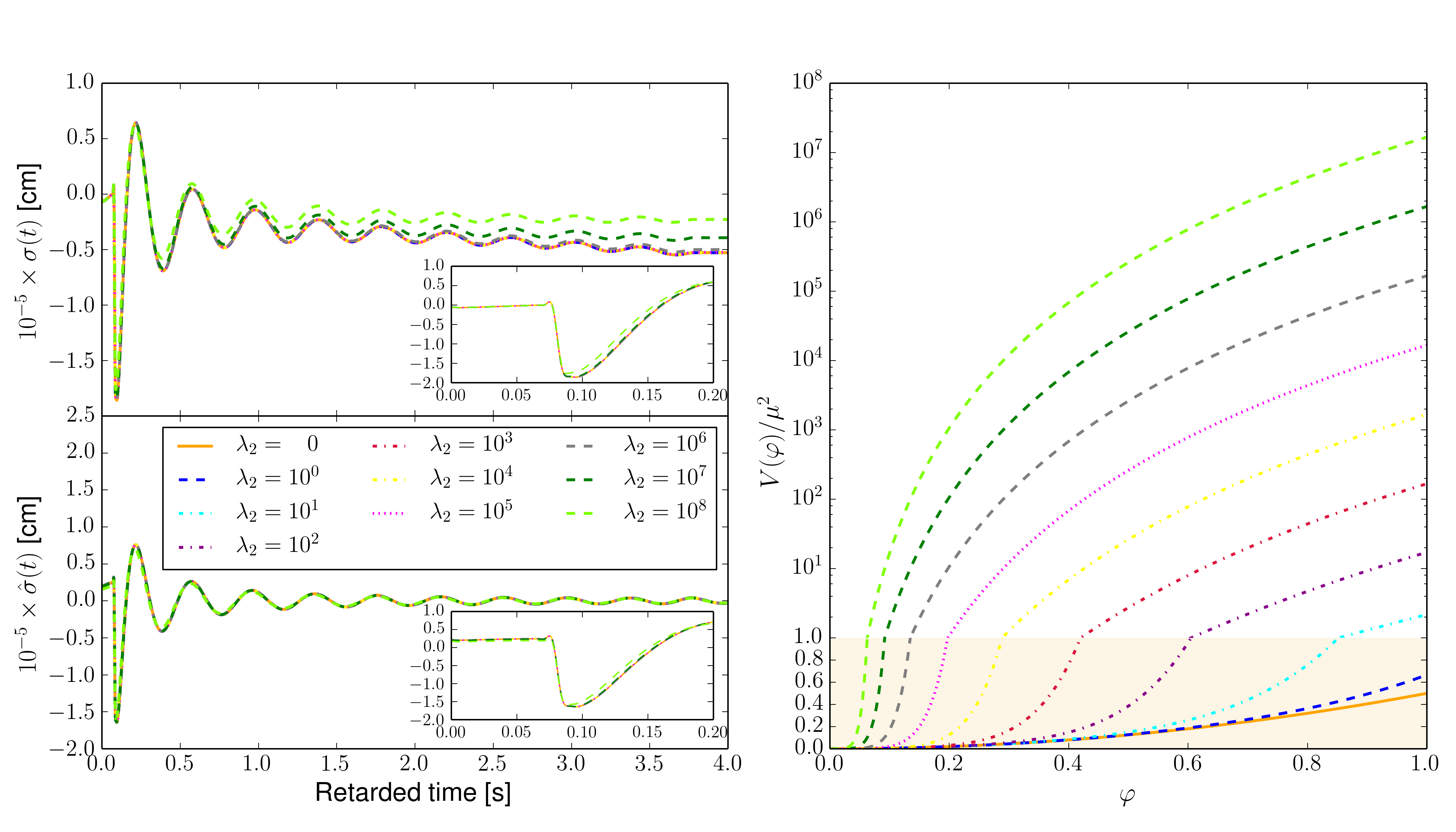}
	\caption{The effect of varying the sextic, $\lambda_2$, term of the scalar field potential. Top-left panel: the scalar field signal at the extraction radius $7\times 10^4$ km. Bottom-left panel: the band-passed scalar field signal at the extraction radius $7\times 10^4$ km; modes with frequencies less than $\omega_*$ are set to $0$. Right-hand panel: the scalar field potential. The progenitor used is a 41 $M_{\odot}$ star of $10^{-4}$ solar metallicity and the scalar parameters are $\alpha_0=10^{-2}$, $\beta_0=-20$.
	%, $\mu=10^{-14}$ eV.
	}
	\label{fig:vary_lambda2}
\end{figure}
\begin{figure}[t]
	\centering
	\includegraphics[width=1.0\textwidth]{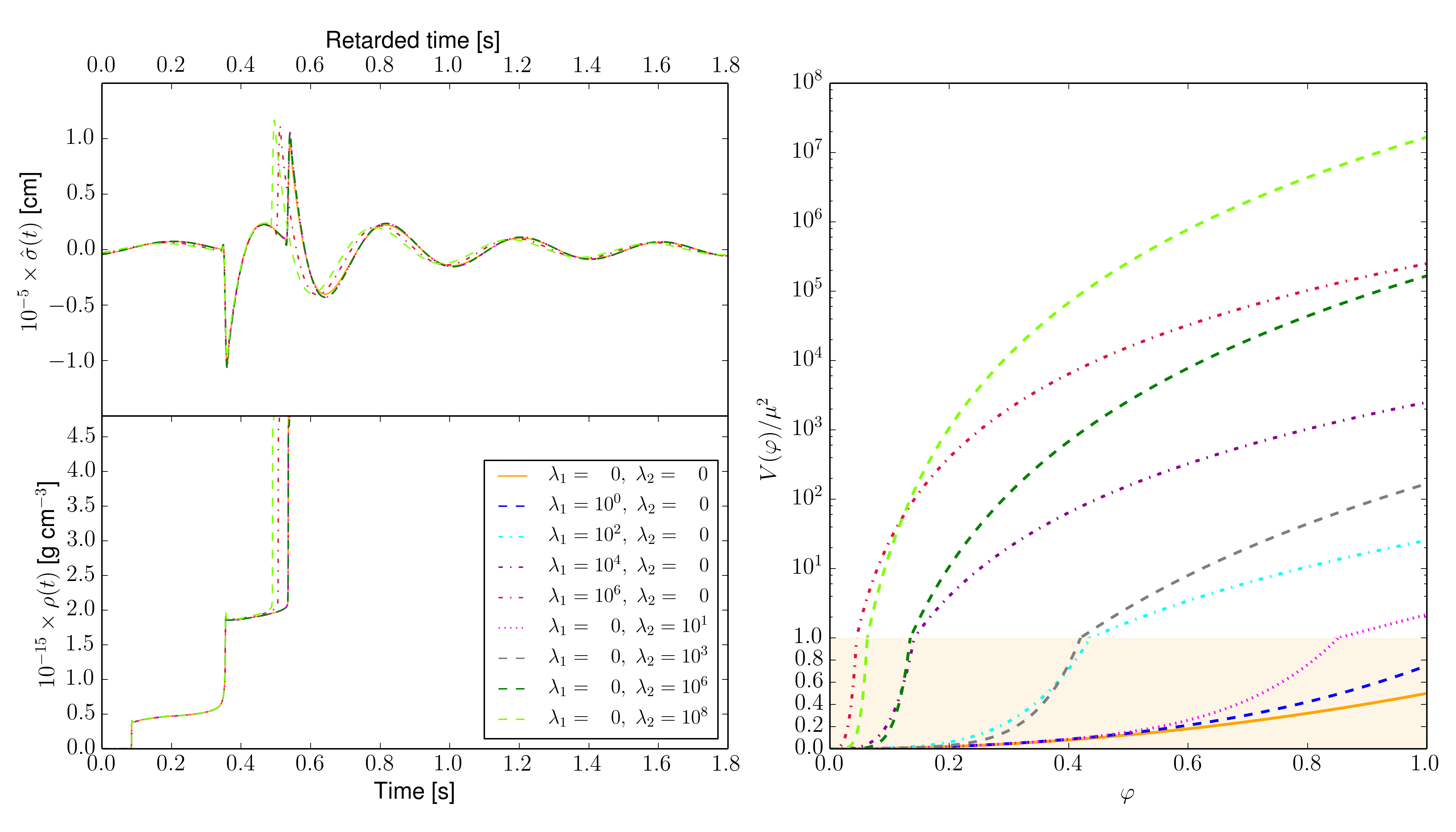}
	\caption{The effect of varying the $\lambda_1$ and $\lambda_2$ parameters for models that lead to BH formation.
	Top-left panel: the band-passed scalar field signal at the extraction radius $7\times 10^4$ km; modes with frequencies less than $\omega_*$ are set to $0$. Bottom-left panel: the central density until a BH is reached. Right-hand panel: the scalar field potential. The progenitor used is a 39 $M_{\odot}$ star of $10^{-4}$ solar metallicity and the scalar parameters are $\alpha_0=10^{-3}$, $\beta_0=-5$.
	%,  $\mu=10^{-14}$ eV.
	}
	\label{fig:BH}
\end{figure}
Next, we repeat this collapse simulation for
non-zero $\lambda_1$ but keeping all $\lambda_i=0$ for
$i \geq 2$. Even though the potential changes dramatically
as we increase $\lambda_1$ (see the right panel
of Fig.~\ref{fig:vary_lambda1}), up to $\lambda_1=10^4$,
we observe no modifications in the scalar wave signals
(see upper left panel of Fig.~\ref{fig:vary_lambda1})
and the peak value of the scalar field at the stellar
centre remains $0.38$. For $\lambda_1 \gtrsim 10^5$,
we start seeing
a mild deviation of the signal in the form of a slow
drift. The low frequency modes
associated with this drift, however, will be screened during the propagation of the signals to large distances: all
modes with frequencies below $\omega_*$
-- 2.42~Hz in this case -- decay exponentially
at large radii \cite{Sperhake:2017itk}. In the context
of LIGO-Virgo observations, these modes
are therefore irrelevant.
In order to asses the observationally relevant impact
of $\lambda_1$, we band pass the signals by suppressing modes below $\omega_*$. The result is shown
in the bottom left panel of Fig.~\ref{fig:vary_lambda1}
and demonstrates excellent agreement of all waveforms
up to $\lambda_1 \approx 10^6$. Increasing
$\lambda_1$ beyond this magnitude leads to a gradual
reduction in the amplitude of the scalar radiation.

In Fig.~\ref{fig:vary_lambda2}, we have repeated the same
analysis but now varying $\lambda_2$ while keeping all
other $\lambda_i=0$. The observations are the same
as for $\lambda_1$ except that the sextic term
only exhibits
visible deviations for $\lambda_2 \gtrsim 10^7$ (instead
of $\lambda_1 \gtrsim 10^5$ for the quartic term). This
quantitative difference is not surprising given
the additional suppression in the sextic term by a factor
$\sim \varphi^2$. 
We have obtained similar results when varying $\lambda_3$:
we observe no significant deviation up to values of
$\lambda_3 \approx 10^{8}$
but a reduction of the
GW amplitudes above this value. For $\lambda_3=10^{12}$, for
example, we see a drop by about $50\%$ relative to the
non-self-interacting case.

The scenario discussed so far represents a straightforward
collapse of a low-density star to a compact neutron star.
Especially for high $M_{\rm ZAMS}$ progenitor models, however,
the outcome can reveal a more complex behaviour. The
collapse may first form a weakly scalarised neutron star
which, through continued accretion, later migrates to the
strongly scalarised branch or the collapse may form a BH
instead of a neutron star. We now consider a model
that exhibits all these features in one collapse.
This model consists of a 39 $M_{\odot}$ progenitor with
$10^{-4}$ solar metallicity collapsing in ST theory with
$\alpha_0=10^{-3}$, $\beta_0=-5$.
For $\lambda_i=0$, this star initially forms a weakly
scalarised compact core, as indicated by the first jump
in the central density at $t\simeq0.09\,\mathrm{s}$
in the bottom left panel of Fig.~\ref{fig:BH}. Through
accretion, the core becomes more massive and eventually
migrates to the strongly scalarised branch at
$t\simeq0.35\,\mathrm{s}$. This change in scalarisation
-- the central value of the scalar field increases from
nearly zero to $0.24$ --
generates the first peak in the scalar wave signal in the upper
left panel of Fig.~\ref{fig:BH}. As the core continues
accreting and becomes yet more massive, ultimately
a BH is formed and the star descalarises
in accordance with the no-hair theorems for BHs
\cite{Hawking:1972qk, Thorne:1971}. From the viewpoint
of GW generation, this
process is, in essence, a reversal of the original
scalarisation and, correspondingly, leads to a second
peak at $t\simeq0.53\,\mathrm{s}$ resembling a mirror image
of the first peak.
This scenario remains largely unchanged
when increasing $\lambda_1$ up to $10^5$
or, alternatively, $\lambda_2$ up to $10^7$.
For larger values, we eventually see a shift in the 
time of collapse to a BH: $t\simeq0.50\,\mathrm{s}$ for 
$\lambda_1=10^6$ and at $t\simeq0.49\,\mathrm{s}$ for 
$\lambda_2=10^8$. 
For even larger values of the quartic and sextic parameters, the 
scalarisation is weakened in accordance with our
observations in the above cases. Furthermore,
the strongly scalarised stage becomes shorter; for
$\lambda_1 = 10^7$ or $\lambda_2 = 10^9$ this stage disappears
and the star collapses directly into a BH at $t\simeq0.35\,\mathrm{s}$. This observation indicates
that strong scalarisation tends to delay BH formation;
we will discuss this phenomenon in more detail
in a follow-up paper \cite{FollowUp}.

%=============================================================================
\section{Self-interaction and wave propagation} 
\label{sec:propagation}
The GW signal we may observe from a stellar collapse event
in massive ST gravity is affected by self-interaction terms
in the potential in two ways: (i) the scalarisation of the
star and the corresponding local generation of a GW signal, and (ii) the propagation of this signal from source to detector.
In the previous section, we have addressed item (i); GW generation is affected by the self-interaction terms only for enormous values of the dimensionless coefficients $\lambda_1$
and $\lambda_2$.
In this section we will discuss item (ii).
We begin by reviewing the case without self-interaction.

In ST theories with $\lambda_i=0$,
the GW signals at astrophysically large distances, $d$, tend to a predictable \emph{inverse chirp}
\cite{Sperhake:2017itk, FollowUp};
at each instant in time
the signal is quasi-monochromatic with instantaneous frequency
\begin{equation}\label{eq:OmegaSPA}
  \Omega(t)=\omega_*/\sqrt{1-\left(d/t\right)^{2}}\quad\textrm{for}\; t>d\,,
\end{equation}
and an amplitude which varies as a function of time according to
\begin{align}
A(t) &= \sqrt{\frac{2}{\pi}} \frac{(\Omega^{2}-\omega_*^{2})^{3/4}}{\omega_* d^{1/2}}\, \Big|\tilde{\sigma}\big[\Omega(t)\big]\Big|\,.\label{eq:SPA}
\end{align}
Here, $\sigma(t)\equiv r_{\rm ex}\varphi(t;r_{\rm ex})$ is the rescaled scalar field signal
extracted at a radius $r_{\rm ex}$ in the wave zone of a
strong-field simulation
(e.g.\,as obtained from our numerical simulations at 
$7\times 10^4\,{\rm km}$),
and a tilde denotes a Fourier transform.
Note that the scalar profile near the source only enters into the expression for the amplitude of the signal at large distances (not the frequency), and even then only through the modulus of its Fourier transform (or power spectrum).

%=============================================================================
\subsection{A toy model} 
The structure of the inverse chirp is due, almost entirely, to the dispersive nature of the wave propagation, and not to any internal dynamics of the neutron star.
To emphasise this point
let us consider the following toy model.

At large distances, the dynamics of the non-self-interacting scalar field are governed by the flat-space Klein-Gordon equation,
\begin{equation} \label{eq:KGequation}
  \left( \partial_{t}^{2} - \nabla^{2} \right) \varphi + \mu^{2}\varphi  = 4\pi\varrho \,,
\end{equation}
where $\varrho$ is the source for the scalar field. 
Although this flat-space equation is only expected to hold at large radii, in our toy model it will be used throughout spacetime.
Before the neutron star has scalarised there is no source term; however, after scalarisation the field is sourced by a small, static neutron star.
We approximate this scenario with a source
function given by
\begin{equation} \label{eq:source_term}
  \varrho(t;\vec{x}) = \varphi_{*}\,W_{\tau}(t)\delta^{(3)}(\vec{x}) \,, \quad \mathrm{where}\;\; W_{\tau}(t) = \begin{cases}
  \frac{2}{\pi}\arctan\left(\frac{t}{\tau}\right) \;&\mathrm{if}\;t\geq 0 \\
  0 \;&\mathrm{if}\;t<0
  \end{cases} \,,
\end{equation}
where $\varphi_{*}(x)$ parameterises the magnitude of the scalarisation, and $\tau$ is the typical timescale over which scalarisation occurs.
The field equation for our toy model, Eq.~(\ref{eq:KGequation}), is linear and may be solved using a retarded Green's function;
\begin{equation} \label{eq:KGsolution}
  \varphi(x) = \int \mathrm{d}^{4}x\;G(x,x')\varrho(x') \,,
\end{equation}
where $G(x,x')$ is given explicitly by \cite{Poisson2004}
\begin{equation} \label{eq:RetardedGreensFunction}
  G(x,x') = \left[ \delta(\chi)-\frac{\mu}{\sqrt{-2\chi}}J_{1}(\mu\sqrt{-2\chi})\theta(-\chi) \right] \quad \mathrm{if} \; t-t' \geq 0 \quad \mathrm{else}\; 0 \,,
\end{equation}
with $\chi(x,x')=\frac{1}{2}\eta_{\mu\nu}(x-x')^{\mu}(x-x')^{\nu}$ and $J_{\alpha}$ denoting the Bessel functions of the first kind. 
Evaluating the integral in Eq.~(\ref{eq:KGsolution}) with the source in Eq.~(\ref{eq:source_term}) gives
\begin{equation} \label{eq:scalar_field_solution}
  \varphi(t;r) =
   \varphi_{*}\Bigg[\frac{W_{\tau}(t\!-\!r)}{r} - \int_{0}^{t-r}\!\mathrm{d}t'\,\frac{\mu J_{1}\big(\mu\sqrt{(t-t')^{2}-r^{2}}\big)W_{\tau}(t')}{\sqrt{(t-t')^{2}-r^{2}}}\Bigg] \quad \mathrm{if} \; t \geq r \quad \mathrm{else}\; 0 .
\end{equation}
This is the solution (written in terms of an integral to be evaluated numerically) for the scalar field in our toy model; the first term is the familiar result for a massless field, whilst the second ``tail'' term depends on $\mu$ and accounts for the dispersive nature of the wave propagation.
\begin{figure}[th]
  \centering
    \includegraphics[width=0.80\textwidth]{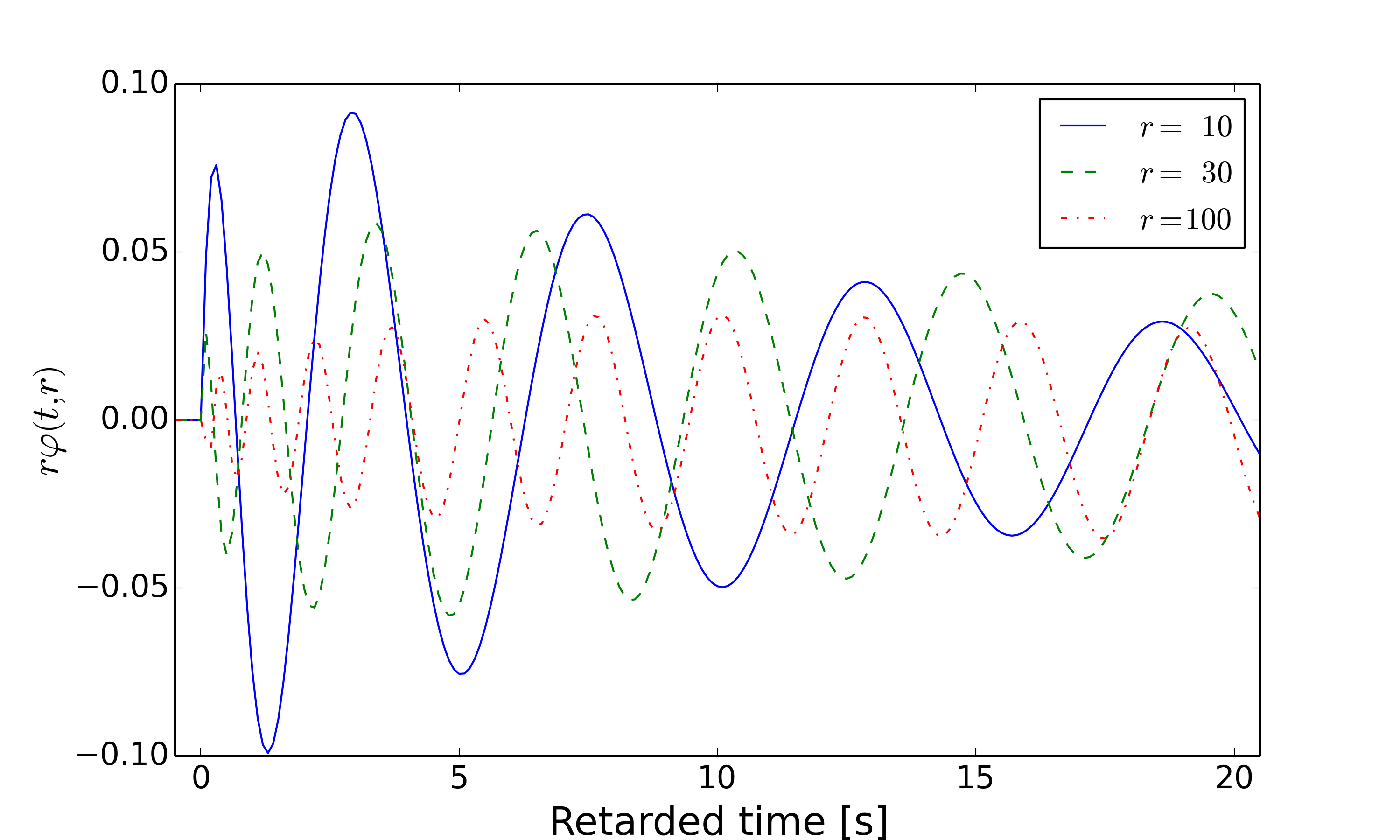}
    \caption{ \label{fig:plot_rphi_toymodel}
    The scalar field as a function of retarded time in the toy model [Eq.~(\ref{eq:scalar_field_solution})] at three different distances from the origin: $r=10,\,30,\;\mathrm{and}\;100$. 
    The other model parameter values are $\mu=\varphi_{*}=\tau=1$, and $\lambda_{i}=0$.
    As described in \cite{Sperhake:2017itk}, the signal becomes increasingly oscillatory at greater distances; the $r=10$ curve has five maxima in the plot range, the $r=30$ curve has seven, whilst the $r=100$ curve has ten.
    }
\end{figure}
The result in Eq.~(\ref{eq:scalar_field_solution}) is plotted as a function of retarded time $t-r$ at several fixed radii in Fig.~\ref{fig:plot_rphi_toymodel}.
Obviously our simple toy model has neglected all of the interesting physics in the region surrounding the scalarising neutron star (for example, it neglects the dynamical space-time curvature, the magneto-hydrodynamics of the collapsing baryons, and the finite size and internal structure of the remnant).
Nevertheless, the scalar profiles plotted in Fig.~\ref{fig:plot_rphi_toymodel} bear a close resemblance to the results of the full numerical simulation reported in Fig.~1 of \cite{Sperhake:2017itk} as well as the profiles for the self-interacting scalar fields in Figs.~\ref{fig:vary_lambda1} and \ref{fig:vary_lambda2}.
The fact that the qualitative features of the true signal can be recovered by a simple toy model based solely on the massive wave equation (\ref{eq:KGsolution}) serves to emphasise the point that, for non-self-interacting scalar fields, the core collapse GW signal observed at large distances is determined almost entirely by the dispersive nature of the wave propagation in flat space-time, and not by the internal dynamics of the GW source.

We now turn our attention to self-interacting scalar fields [i.e.\ $n>0$ in Eq.~(\ref{eq:Def_Potential})].
The above analysis of our toy model cannot be repeated for this case because the flat space wave equation is now non-linear and does not admit a solution using Green's functions;
\begin{equation} \label{eq:KGequation_nonlinear}
  \left( \partial_{t}^{2} - \nabla^{2} \right) \varphi + \mu^{2}\varphi \left(1+\lambda_1\varphi^2+\lambda_2\varphi^4+\ldots+\lambda_n\varphi^{2n}\right)  = 4\pi\varrho \,.
\end{equation}
However, the non-linear terms in this equation are only significant near the origin where $\lambda_k\varphi^{2k}\sim 1$; 
at large distances the scalar field decays as $\varphi\lesssim 1/r$ (or faster).
Therefore, the propagation of the signal from the origin to some intermediate radius $r_{\rm ex}$ where $\lambda_1\varphi^{2}\ll 1$ is governed by the full non-linear equation. 
However, the subsequent propagation from $r_{\rm ex}$ out to the astrophysical distances $d$ relevant for LIGO/Virgo observations is once again governed, to an excellent approximation, by the linear Eq.~(\ref{eq:KGequation}). 
Therefore, we expect that at very large distances the signal will still be an inverse chirp as described by Eqs.~(\ref{eq:OmegaSPA}) and (\ref{eq:SPA}).

%=============================================================================
\subsection{Numerical integration of the non-linear Klein-Gordon equation}
The expectation from the above discussion is that
higher-order effects in the potential should not
significantly affect the propagation of the gravitational
wave signal provided a sufficiently large $r_{\rm ex}$ is
chosen.
We test this hypothesis by numerically solving
the three dimensional Klein-Gordon equation
(\ref{eq:KGequation_nonlinear}) with $\varrho=0$; here we do
not source a signal through some model function
$\varrho$, but instead inject a genuine core collapse
waveform at $r=r_{\rm ex}$ and propagate it
further outwards to $r=d$
with the homogeneous Klein-Gordon equation.
Writing this equation in terms of
the variable $\sigma=r\varphi$ (and restoring factors of
$\hbar$), we obtain a
one-dimensional wave equation that we evolve in characteristic
coordinates $(u:=t-r,r)$ in the form of the first-order system
\begin{eqnarray}
  \partial_r \sigma &=& \eta\,,
  ~~~~~~~~~~\varphi=\frac{\sigma}{r}\,,
  \nonumber \\[5pt]
  \partial_u \eta &=& \frac{1}{2}\partial_r \eta-
      \frac{1}{2}\sigma \frac{\mu^2}{\hbar^2}
      \left(1+\lambda_1\varphi^2+\lambda_2\varphi^4+\ldots+
      \lambda_n \varphi^{2n}\right)\,.
      \label{eq:KGchar}
\end{eqnarray}
The specific injected signal in our case is
given by the waveform $\sigma(t,r_{\rm ex})$ extracted
at $r_{\rm ex}=7\times 10^4\,{\rm km}$
from the
collapse of a $12\,M_{\odot}$ progenitor in non-self-interacting ST gravity with $\alpha_0=0.01$, $\beta_0=-17$.
This is a representative example of a waveform
generated for a strongly scalarising star, but it here only serves
as a common starting signal for the Klein-Gordon
evolutions; what we are
interested in is how the propagation of this signal from
$r_{\rm ex}$ to $d$
changes for different choices of the $\lambda_i$
while setting all other $\lambda_i=0$ and
fixing $\mu=10^{-14}\,{\rm eV}$.

\begin{figure}
    \centering
    \includegraphics[width=0.9\textwidth]{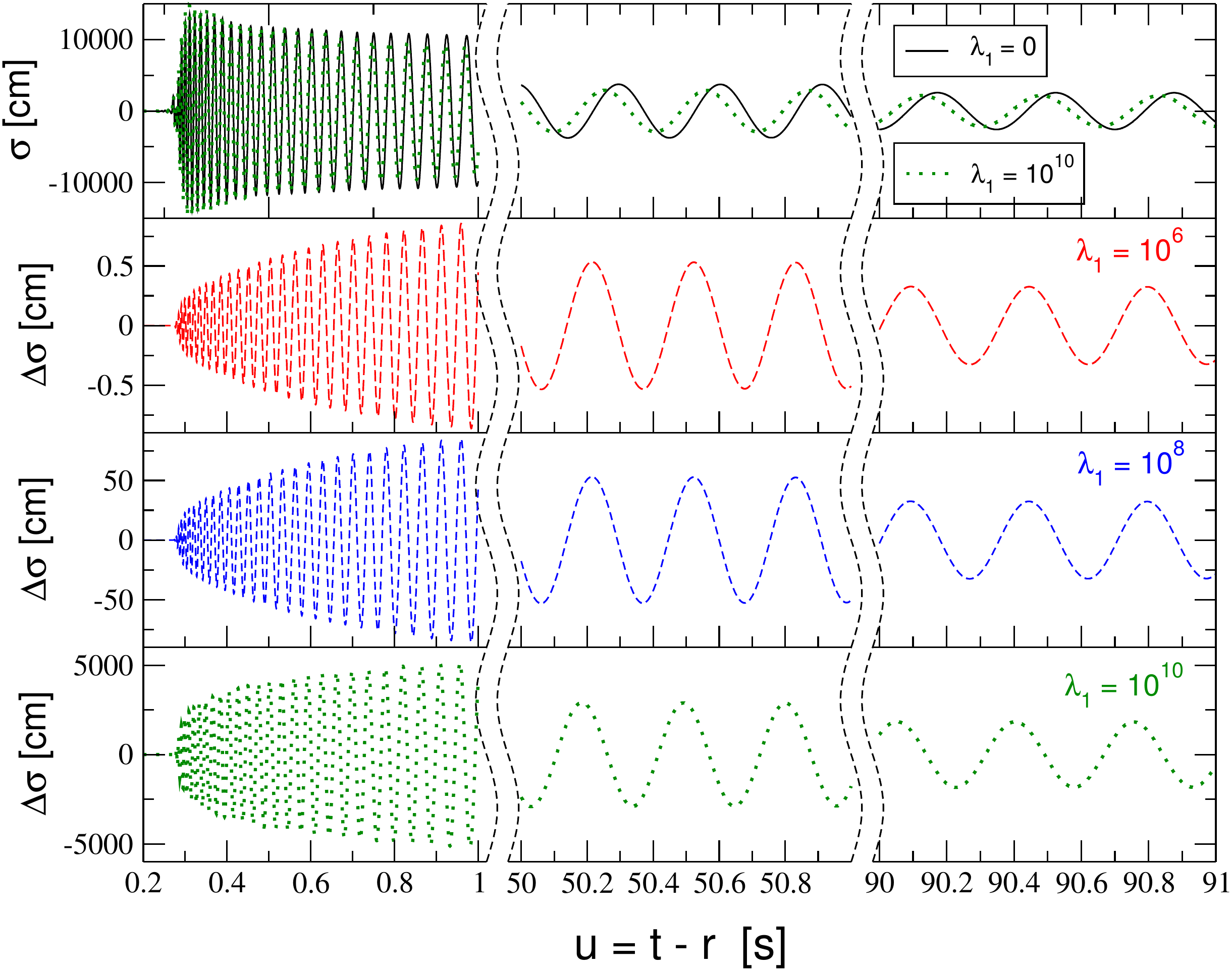}
    \caption{A representative scalar wave signal is injected at
    $r_{\rm ex}=7\times 10^{4}\,{\rm km}$ and then propagated
    according to Eq.~(\ref{eq:KGchar}) to
    $d=3\times 10^{7}\,{\rm km}$
    for $\lambda_1=0$, $10^6$, $10^8$ and $10^{10}$, respectively.
    The top row shows the signals thus obtained at 
    $d =3\times 10^{7}\,{\rm km}$;
    the waveforms for $\lambda_1=10^6$ and $\lambda_1=10^8$ would
    be indistinguishable from the $\lambda_1=0$ curve and are
    not included to avoid confusion. Note that it would be impossible
    to resolve the long and highly oscillatory signal
    over its entire duration in this
    figure and we instead display 3 segments of length 1\,s each
    around u=0,\,50\, and 90\,s. The three following rows respectively
    show the difference of the three signals propagated with
    self-interacting potential from the $\lambda_1=0$ case.
    Note that the deviation grows linearly with $\lambda_1$
    and is negligible up to about $\lambda_1=10^8$.}
    \label{fig:propagation100s}
\end{figure}

The signal propagated to $d=3\times 10^7\,{\rm km}$
with $\lambda_1=0$ and
$\lambda=10^{10}$, respectively, is shown by the two curves in
the top row of Fig.~\ref{fig:propagation100s}.
For $\lambda_1=10^6$ and $\lambda_1=10^8$ the curves would be
indistinguishable from the $\lambda_1=0$ case in this plot and
we therefore do not include them. The differences in the
propagated signals for all self-interacting cases relative to the
$\lambda_1=0$ signal are shown in the three remaining rows
of Fig.~\ref{fig:propagation100s}; note that this deviation
increases linearly with $\lambda_1$ but remains well below the
percent level even for $\lambda_1=10^8$.

Even within the relatively simple
framework of the characteristic
Klein-Gordon equation, it is not feasible to compute the propagation
of the signal to distances of the order of 10\,kpc, typical
for galactic LIGO sources.
We have, however,
repeated the above numerical analysis for
$d=6\times 10^{7}\,{\rm km}$ to assess how our findings
change with extraction radius. We find no significant
difference; the analogous plot for $d=6\times 10^{7}\,{\rm km}$  is barely distinguishable from Fig.~\ref{fig:propagation100s}.
In conclusion, the propagation of the wave signal in
ST theory with higher-order
self-interaction terms remains essentially unchanged
relative to the massive case studied in Ref.~\cite{Sperhake:2017itk},
provided $\lambda_1 \lesssim 10^8$. For larger values of $\lambda_1$,
significant quantitative deviations arise but do not change the ``inverse-chirp''
character nor the oscillatory pattern of the signal. This
result is in agreement with a rough estimate of the accumulation
of deviations in the self-interacting case: Over a distance $d$,
the leading extra terms in the potential are of order
$\mathcal{O}(d^{-2})$ and the total accumulated
deviation is dominated by the contributions at {\em small} distance
from the source. From 
Figs.~\ref{fig:vary_lambda1}-\ref{fig:BH}, we see that the scalar field
reaches values of the order of $10^{-5}$ at
$r_{\rm ex}=7\times 10^{4}\,{\rm km}$, so that
$\lambda_1 \varphi^2 \sim 1$ for $\lambda_1=10^{10}$, which
is precisely the magnitude of $\lambda_1$ where
deviations in the scalar wave propagation become significant.

For the sixth-order and higher terms in the potential, the same
argument holds even more emphatically: The $\lambda_2$ term in the
potential falls of with $d^{-4}$ and has even less impact on the
wave propagation than the quartic terms and likewise for $\lambda_3$
etc.

%=============================================================================
\section{Conclusions}
\label{sec:conclusions}
Scalar-tensor theories are perhaps the theoretically most well understood of all possible modifications to the general theory of relativity.  
In such theories compact objects, such as neutron stars, can undergo {\em spontaneous scalarisation} \cite{Damour:1993hw}, where the scalar field inside the star grows rapidly to large values of order $\mathcal{O}(1)$.
This process, and the accompanying GW signal produced, has been studied dynamically during the core collapse process by a number of authors \cite{Scheel:1994yr,Harada:1996wt,Novak:1997hw,Novak:1998rk,Novak:1999jg,Palenzuela:2013hsa,Palenzuela:2015ima,Gerosa:2016fri}.

If the scalar is endowed with a mass (i.e.\ a quadratic potential)
a much wider range of the theory's parameter space is compatible
with present observational constraint, allowing for a significantly
enhanced magnitude of the scalarisation, and the GW signal gets distorted by the effects of dispersion and tends to a universal {\em inverse-chirp} profile \cite{Sperhake:2017itk}.

In this article we have investigated the effects of self-interaction in the scalar field (i.e.\ a more complicated potential including several higher-order terms).
Our results should be compared with those of \cite{Cheong:2018gzn} where the case of the
quartic potential was studied and it was found that self-interaction can
suppress the scalarisation effect. Our results are consistent with this
conclusion, but we note that such
suppression requires a great
deal of fine-tuning of the parameters of the potential.
For example, when including a quartic $\varphi^{4}$ term, the
dimensionless coefficient must be artificially set to a factor of
$\gtrsim 10^{5}$ larger than the leading quadratic term before there is
any noticeable effect (see Fig.~\ref{fig:vary_lambda1}).
The main conclusion of our study is that the scalarisation 
phenomenon is incredibly robust against
the effects of scalar self-interaction.

We have also demonstrated that the important features of the wave propagation are unaffected by the presence of self-interaction in the scalar potential. 
In particular, the observed signal still tends to the same universal inverse-chirp profile at large distances.

This apparent insensitivity to the detailed shape of the potential may be both a blessing and a curse.  
On the one hand it allows a very simple and robust prediction of the expected GW signal to be made.
However, if such a signal should ever be detected it will make it very difficult to observationally study any details of physics in the scalar sector, beyond measuring the scalar field mass. 

%=============================================================================
\section*{Acknowledgments}
We thank Davide Gerosa, David Hilditch, and Christian Ott
for helpful discussions.
This work was supported by
the European Union's H2020 ERC Consolidator Grant
``Matter and strong-field gravity: New frontiers in Einstein's
theory'' grant agreement no. MaGRaTh--646597 funding from the
European Union's Horizon 2020 research and innovation programme
under the Marie Sk\l odowska-Curie grant agreement No 690904, the
COST Action Grant No.~CA16104, from STFC Consolidator Grant No.
ST/P000673/1, the SDSC Comet and TACC Stampede2 clusters through
NSF-XSEDE Award Nos.~PHY-090003,
and Cambridge's CSD3 system
system through
STFC capital grants ST/P002307/1 and ST/R002452/1 and STFC operations grant ST/R00689X/1.
R.R.-M. acknowledges support by a STFC studentship.
%=============================================================================
\section*{References}
%
%\bibliography{newuli}
\bibliographystyle{unsrt}

\end{document}